# Enhancing Automated and Early Detection of Alzheimer's Disease Using Out-Of-Distribution Detection

**Audrey Paleczny[1], Shubham Parab[2], and Maxwell Zhang[3]**

[1]Woodside Priory High School, fromaudreytoyou@gmail.com
[2]Lynbrook High School, skparab1@gmail.com
[3]Palo Alto High School, maxwell.zhang623@gmail.com

*all authors contributed equally

### ABSTRACT

More than 10.7% of people aged 65 and older are affected by Alzheimer's disease. Early diagnosis and treatment are crucial as most Alzheimer's patients are unaware of having it until the effects become detrimental. AI has been known to use magnetic resonance imaging (MRI) to diagnose Alzheimer's. However, models which produce low rates of false diagnoses are critical to prevent unnecessary treatments. Thus, we trained supervised Random Forest models with segmented brain volumes and Convolutional Neural Network (CNN) outputs to classify different Alzheimer's stages. We then applied out-of-distribution (OOD) detection to the CNN model, enabling it to report OOD if misclassification is likely, thereby reducing false diagnoses. With an accuracy of 98% for detection and 95% for classification, our model based on CNN results outperformed our segmented volume model, which had detection and classification accuracies of 93% and 87%, respectively. Applying OOD detection to the CNN model enabled it to flag brain tumor images as OOD with 96% accuracy and minimal overall accuracy reduction. By using OOD detection to enhance the reliability of MRI classification using CNNs, we lowered the rate of false positives and eliminated a significant disadvantage of using Machine Learning models for healthcare tasks. Source code available upon request.

**Keywords:** Alzheimer's disease, *convolutional neural network* (*CNN*), random forest, out-of-distribution (OOD), supervised learning, magnetic resonance imaging (MRI), brain scan

### I. INTRODUCTION

With almost 1 in 9 people aged 65 and older suffering from Alzheimer's, there is a pressing need for earlier and more accurate disease detection (Texas DSHS, Rasmussen et al.). Researchers have explored the potential of using brain region volume to detect Alzheimer's and have observed promising results (Thompson et al., Brewer et al.). A study by S Lehéricy et al. demonstrated that hippocampal and amygdala volume could be used as indicators of Alzheimer's disease. They recorded a 40% decline in amygdala volume, 45% decline in hippocampal volume, and 63% ventricle enlargement for moderate Alzheimer's patients compared to control group participants (Lehericy et al.). In addition to volume, region thickness was also affected, which a study by Mert R. Sabuncu et al. explored. This study examined cortical and hippocampal thicknesses using FreeSurfer, a neuroimaging analysis software (Reuter et al.). The researchers concluded that baseline thicknesses of areas vulnerable to Alzheimer's were significantly smaller than the presymptomatic group (Sabuncu et al.). In this case, baseline thickness refers to the original region thickness for patients when they started the study. Building on these observations with technological tools, a paper by Walter H. L. Pinaya et al. detailed how normative modeling can be used to identify critical regions affected by Alzheimer's and computed deviations that correlated with the severity of the condition (Pinaya et al.). After training an encoder on a dataset of healthy controls, the researchers estimated how patients deviated from the controls and what caused this deviation. This new methodology matched the cross-cohort generalizability of traditional classifiers, performing well when applied to other datasets.

Machine learning has also been explored in this field. In a study by Afreen Khan et al., researchers use a 5-stage machine learning pipeline with the OASIS (Open Access Series of Imaging Studies) dataset to classify MRI brain scans of 150 subjects into Alzheimer's sub-levels (Khan et al.). By using a Random Forest ensemble classifier, researchers can take advantage of this workflow's individuality to create a model with 86.84% classification accuracy. Going beyond feature-based models, researchers have also experimented with Convolutional Neural Networks (CNNs). In a study by Shakila Shojaei et al., researchers used MRI brain scans from the Alzheimer's Disease Neuroimaging Initiative (ADNI) dataset. They trained a 3D CNN model with a brain mask for Alzheimer's disease patients (Shojaei et al.). With 5-fold cross-validation, their model achieved a 93% validation accuracy with 29 brain regions using the lrp_z_plus_fast explainability method. In another study, researchers have further shown the efficacy of CNNs in detecting the presence of Alzheimer's disease. Karl Bäckström et al. created a CNN to detect MRI brain scans for Alzheimer's. By optimizing hyperparameters, batch size, and amount of preprocessing applied to images, the researchers created a model to detect the scans for Alzheimer's with 98.74% accuracy (Bäckström et al.).

However, while these models show excellent overall performance, they can produce false positives when they are subject to images without Alzheimer's, which can lead to unnecessary treatment (Gaugler et al.). These false positives may lead to further complications, including high costs and potential harm to the patient (Hunter et al., Happich et al.). These false positives make such models unreliable, so finding a method to reduce these misdiagnoses without impacting the overall accuracy is vital. In addition, a significant portion of the high-performing models, especially the 3D CNN, used by past research require massive computational resources, which are not easily accessible to physicians.

To overcome these limitations, we reduce the number of false positives generated by the model by implementing out-of-distribution (OOD) detection in our 2D CNN. This helps identify results likely to be incorrect predictions, flagging them as "unsure" and allowing for human review. We experimented with brain volume-based Random Forest models and 2D CNN output-based Random Forest (stacked ensemble) models. We then integrated OOD detection into our 2D CNN by splitting 3D images into layers and training the CNN on the individual layers. This lowered the processing power required to train the model. Beyond these novelties, our CNN output-based Random Forest model included unique data selection for training to limit model confusion. Moreover, our model's maximum number of filter channels is substantially lower than current state-of-the-art research, which allows our model to be optimized on lower-end processing units while still maintaining high accuracy. We only use a maximum of 4 channels, while other researchers utilize up to 1920 channels (Mujahid et al.) or even 2048 channels (Kumar and Sasikala). Our CNN output-based Random Forest model outperformed our preprocessed image volume-based model with an accuracy of 98% compared to 93% for binary classification and an accuracy of 95% as opposed to 87% for 3-class classification. After applying the OOD detection to the CNN model, we could flag 96% of the OOD dataset as OOD, with only a 4% decline in overall model accuracy.

By creating lower computational power models to produce results comparable to methods requiring much more power, we make Alzheimer's detection using MRIs more easily accessible to people who may not have such resources. By lowering the false positive result rate and increasing the reliability of this methodology, we take a step towards the real-world usability of machine learning as a more trustable diagnostic tool for diseases.

## II. METHODS

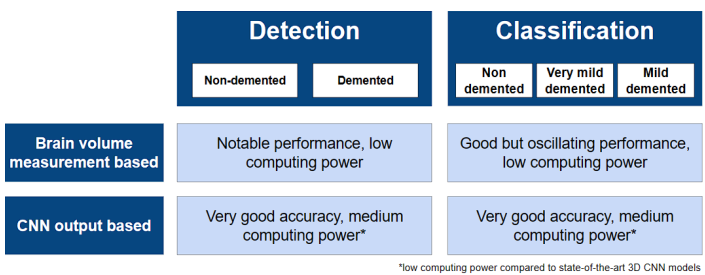

**FIG. 1:** Objectives and observations when training models

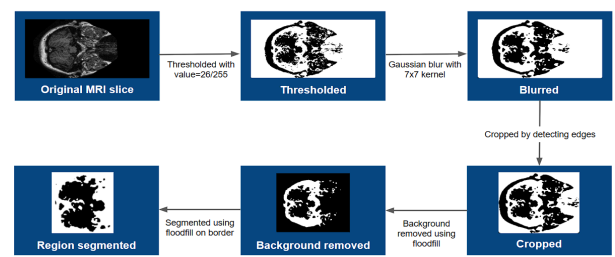

**FIG. 2.** Methodology for brain-volume calculation, feature extraction, and model training

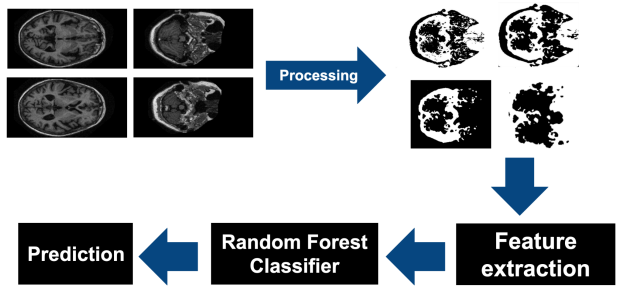

**FIG. 3.** Methodology for brain-volume calculation, feature extraction, and model training

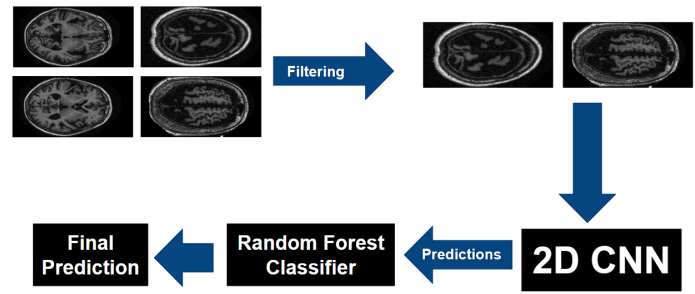

**FIG. 4.** Methodology for CNN output based Random Forest model

| Layer | Input | Param/Kernel | Stride | Padding | Activation | Output |
|---|---|---|---|---|---|---|
| Input | 248 x 248 x 1 | - | - | - | - | 248 x 248 x 1 |
| Conv2D | 248 x 248 x 1 | 5 x 5 | 1 | 1 | ReLU | 246 x 246 x 2 |
| BatchNorm2D | 246 x 246 x 2 | - | - | - | - | 246 x 246 x 2 |
| Conv2D | 246 x 246 x 2 | 5 x 5 | 1 | 1 | ReLU | 244 x 244 x 4 |
| BatchNorm2D | 244 x 244 x 4 | - | - | - | - | 244 x 244 x 4 |
| MaxPool2D | 244 x 244 x 4 | 2 x 2 | 2 | 0 | - | 122 x 122 x 4 |
| Flatten | 122 x 122 x 4 | - | - | - | - | 59536 x 1 |
| Linear | 59536 x 1 | - | - | - | ReLU | 32 x 1 |
| Dropout | 32 x 1 | 0.3 | - | - | - | 32 x 1 |
| BatchNorm1D | 32 x 1 | - | - | - | - | 32 x 1 |
| Linear | 32 x 1 | - | - | - | ReLU | 3 x 1 |
| BatchNorm1D | 3 x 1 | - | - | - | - | 3 x 1 |
| SoftMax | 3 x 1 | - | - | - | - | 3 x 1 |

**Table 1.** Layers of the 2D CNN Model

**MRI Scan Dataset**

We used the Open Access Series of Imaging Studies (OASIS) large-scale brain MRI dataset for deep neural network analysis for Alzheimer's detection (Aithal), which contained 80,000 images. The dataset contained MRI scans from 1417 sessions involving 461 different patients. The 3D MRI brain scans were sliced along the z-axis to produce 256 layers, and we used layers 100 to 160 of each MRI scan. We accounted for these specifications when conducting data processing. However, since our dataset contains moderate dementia examples for only two patients, this was insufficient to create a category. Thus, we excluded this category from our classification models.

**Methodology overview**

We addressed both detection and stage-based classification of images for Alzheimer's Disease. After processing images to segment areas of importance and computing these volumes, we trained a low-powered Random Forest model. We also trained a CNN on raw images from our dataset and used the CNN output data to train a Random Forest model, which performed with high accuracy. We then applied OOD detection to the CNN to flag OOD images.

**Brain scan volume-based Random Forest**

*1. Data Usage*

Our MRI scan dataset was heavily skewed towards "non-demented" images. Therefore, to create a balance between total "demented" and "non-demented" patient scans used, we omitted a number of "non-demented" scans from our dataset. The total number of MRI scans with very mild, mild, or moderate dementia was 311. Based on this measurement, we used only 308 scans from our "non-demented" dataset. This minimized the skew between categories and helped us optimize the data and produce meaningful results. Similarly, for classification, each of the "demented" categories contained far fewer images than the "non-demented" category. Due to this, we again used a limited amount of non-demented MRI scans, which allowed us to obtain a 308:225:82 data split between "non-demented," "very mild demented," and "mild demented" scans, respectively.

*2. Image preprocessing*

We used five features from the images to train our model to create a model with low enough processing power to run on a typical personal computer within a minute on a CPU. These features were extracted from images and included total brain area, area of cerebrospinal fluid, brain area after segmentation, cropped image height, and cropped image weight. As shown in Figure 1, an original MRI slice was taken and thresholded, and a Gaussian blur was applied to remove noise. The edges were detected by scanning the blurred image and cropping off all columns until a color change was observed. At this point, the first feature, the total brain area, was extracted. Subsequently, the background was removed by flood fill from all four corners of the image and the midway edge points. The second feature, the area taken by the cerebrospinal fluid, was measured at this stage. Cerebrospinal fluid is the fluid that takes up the volume that a brain loses when it shrinks due to Alzheimer's. Following this, a second iteration of flood fill was employed to segment the target brain region, and the area was recorded. The height and width of the cropped images were also stored so that the image's dimensions could be considered when normalizing image data for the model. Specifically it was more important to consider the percentage volume of specific regions instead of the areas in pixels, which likely depends on overall brain size in addition to Alzheimer's. This process was repeated for 61 layers of the MRI scan for each patient, totaling 38,280 images. Images were read from the source folder, processed, measured for features, and written into a processed file directory for future reference. The five measured features were stored in a CSV file along with Alzheimer's stage and an identification number that could be used to trace the features back to the MRI scan used.

*3. Model training*

As shown in Figure 2, we trained a random forest model for two tasks. Detection, which required the model to determine whether a specific scan had any level of dementia or was non-demented, was done by training a random forest classifier on 619 sets of features with a 311:308 data split. In this case, the "very mild demented," "mild demented," and "moderate demented" images are all classified in the "demented" category. We also trained a random forest classification model to classify images into "non-demented," "very mild demented," and "mild demented" categories, for which our data had a 308:225:82 split. As previously stated, the "moderate dementia" category was excluded from classification due to its limited number of scans. Both models were fed the same set of 3 features: total brain volume after thresholding, total brain volume after segmentation, and cerebrospinal fluid volume after segmentation, all with respect to image area. These models were trained with a train-to-test split of 80:20 based on patients, and the f-score was calculated with a weighted average. Since Random Forest models internally validate their performance using out-of-bag evaluation, a validation set was not used.

**CNN output based Random Forest**

*1. Data usage for Classification*

To train our classification CNN, we used three subsets of our dataset: "non-demented," "very mild demented," and "mild demented." We split this data into 60%, 20%, and 20% for the train, validation, and test subsets. The splitting was done by a patient rather than a scan, ensuring that all MRI scans associated with a person are within the same category. We adopted this approach because each patient underwent up to 4 MRI scans. Subsequently, we utilized 61 of the 256 2D cross-sectional images per MRI scan in the dataset.

*2. Data augmentation*

We decided to augment our training images to allow our CNN to generalize to images presented in different ways and be more robust. This would allow us to account for different brain sizes among patients and allow models to see more variation than the streamlined data that our dataset provided. To augment images, we applied a random zoom, using RandomAffine to scale it between 80% and 120%. Then we applied a random horizontal and vertical flip, followed by a resize to 248x248 to keep the model input image dimensions constant.

*3. Solely using a 2D CNN*

We designed a 2D CNN Model with a limited number of parameters, as shown in Table 1, in order to minimize training time on low-end GPUs. We then trained this model and evaluated its performance by comparing the predictions to their ground truths. Initially, we added all available images into the datasets and predicted Alzheimer's based on individual slices. We measured metrics like accuracy, precision, recall, and f-score, but they all reported extraordinarily low values. We reviewed our dataset and noticed that Alzheimer's Disease affected varying portions of the brain. Therefore, our consideration of just a single slice of the brain was insufficient to determine the severity of Alzheimer's Disease, and we needed all 61 slices.

Furthermore, we needed to train the model on slices that contained the given severity of Alzheimer's associated with the patient to not throw off the model with confusing data. To do this, we removed brain slices categorized as "mild demented" and "very mild demented" that looked normal in the training and validation dataset. This enabled the validation accuracy to precisely represent the actual accuracy of the model in detecting demented brain slices.

*4. Filtering useful data*

To identify images that accurately represented certain levels of Alzheimer's disease, we processed each image by computing the number of pixels within the skull that represent the cerebrospinal fluid that replaces the volume of a brain when it shrinks. This method was used as shrunken brain size is a primary indicator of Alzheimer's. To do this, we removed the black pixels surrounding the brain. Subsequently, we enhanced the contrast ratio by multiplying the image matrix by 8. We applied thresholding to the image to extract all of the black pixels and calculated the percentage of the total skull area that these pixels occupied. This step enabled us to determine the extent of brain loss in a given slice. We then set a minimum threshold for brain loss, which allowed us to effectively rule out slices of the brain that looked normal but were classified as Alzheimer's, removing a potentially major problem that threw off our model. This filtration was vital as it allowed the model to correctly classify demented images, as the model would get confused if it observed similar patterns across multiple categories. However, we left the test dataset untouched since we wanted the model to output a prediction for every layer of a given brain.

*5. Using CNN with modified data and Random Forest (Stacked Ensemble)*

As shown in Figure 3, we retrained our 2D CNN on the modified data. For each scan, we stored an array containing the number of each label that the CNN predicted for each scan. We also labeled the arrays with the ground truth severity of Alzheimer's disease. We stored these arrays for each scan in a CSV file and trained a Random Forest Classifier on the data with a 70% and 30% split for the test and train datasets, respectively. After testing the model several times to obtain peak performance, we calculated the accuracy and f-score, which were significantly higher than our volume-based model. For the detection of Alzheimer's, a similar process was used, but the "very mild demented" and "mild demented" categories were combined into a single "demented" category, to create two categories: "demented" and "non-demented." We stored the prediction data from our 2D CNN into a .csv file and applied a Random Forest classifier on the data, also using a 70% and 30% split for the train and test datasets, respectively. This model, too, had a much higher accuracy and f-score than our volume-based classifier.

**OOD Detection Implementation**

While training a Random Forest on CNN outputs enhanced its accuracy and overall performance, adding OOD detection would be difficult. This is because the Random Forest confidence levels do not accurately reflect the confidence levels of the CNN, as the Random Forest is the second model in the process, which complicates evaluating the true confidence. Therefore, we excluded the Random Forest model and only used the 2D CNN model for implementing OOD detection. After six epochs, the model performed with a validation accuracy of 88.52% and a test accuracy of 87.27%. We tested this model on "demented" and "non-demented" images, as well as an OOD dataset containing glioma brain tumors found in the Brain Tumor Segmentation 2022 Challenge dataset (Abdallah), which are OOD for our model. By measuring and averaging the confidence levels outputted by predictions for both in-distribution and OOD datasets, we observed that the prediction confidence for OOD images averaged around 56% and rarely surpassed 60%. In contrast, predictions for in-distribution images had confidence averaging 67%. We thus established 60% as an optimal cutoff between OOD and in-distribution images, under which the input image was likely to be OOD. We then integrated this cutoff into our prediction function for test data and configured it to output "unsure" if the confidence was under the threshold. This modified the model so that a significant portion of the OOD data would be flagged. In addition, to test this methodology's effect on in-distribution images, we tested it with our primary dataset.

## III. RESULTS

| Model | Task | Accuracy | F-score |
|---|---|---|---|
| Brain Volume RFC | Detection | 93% | 93% |
| CNN RFC | Detection | 98% | 99% |
| Brain Volume RFC | Classification | 87% | 87% |
| CNN RFC | Classification | 95% | 94% |

**Table 2:** Comparison of performances of brain volume-based and CNN output-based Random Forest Classifier (RFC) for detection and classification.

| Model | Overall Accuracy | OOD Samples Flagged |
|---|---|---|
| Regular Detection CNN (Without OOD Detection) | 88% | - |
| Detection CNN with OOD Detection | 84% | 96% |

**Table 3:** Comparison of CNN performance with and without OOD detection.

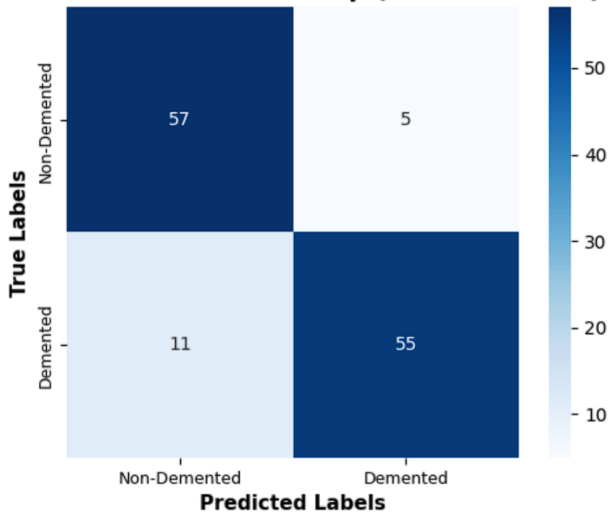

**FIG. 5.** Confusion matrix for brain volume-based Random Forest binary classifier

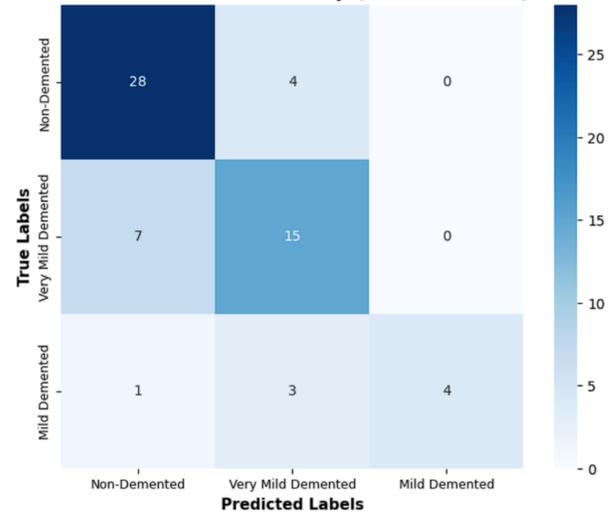

**FIG. 6.** Confusion matrix for brain volume-based Random Forest 3-class classifier

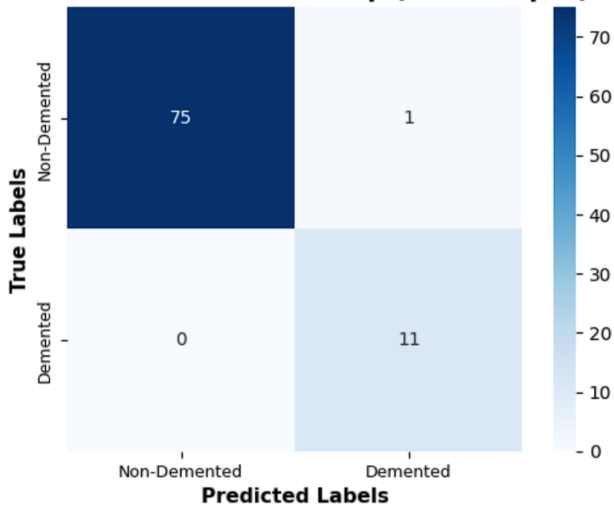

**FIG. 7.** Confusion matrix for CNN output-based Random Forest binary classifier

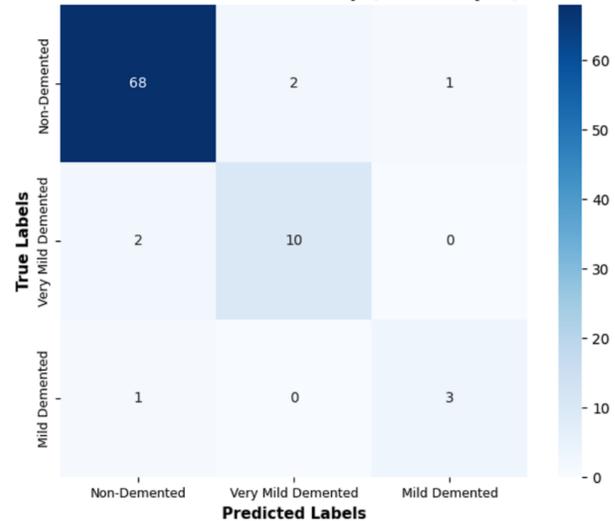

**FIG. 8.** Confusion matrix for CNN output-based Random Forest 3-class classifier

*1. Volume based Random Forest*

Our model trained on volume data was trained 20 times with different test/train subsets. For Alzheimer's detection, the model performed with a peak accuracy of 93% and an average accuracy of 85%. The peak detection f-score was 93%, and the average f-score was 85%. For classification, the peak accuracy was 87%, and the average accuracy was 80%. The peak classification f-score was also 87%, with the mean being 80%. As the confusion matrices in Figures 5 and 6 show, the predictions broadly align with the ground truths. While both classes had similar dataset sizes for detection, the classification data was skewed towards the "very mild demented" and "non-demented" classes. Nonetheless, the model still predicted Alzheimer's severity with notable accuracy.

*2. CNN output based Random Forest (Stacked Ensemble)*

Our second model, a stacked ensemble built by inputting cumulative CNN classification predictions into a Random Forest model, performed with 98% detection accuracy for classifying images into the "demented" and "non-demented" classes, as shown in figure 7. The model also performed with 95% accuracy for 3-class classification, between the "non-demented," "very mild demented," and "mild demented" classes, as shown in figure 8. This performance places our methodology on par with leading studies in the area. In addition, it completes the task with far less processing power.

*3. OOD enhancement*

Integrating OOD detection into our CNN model significantly enhanced our model's performance on the OOD dataset. It was able to flag 96% of our OOD data as OOD, allowing for a significant accuracy boost for the OOD dataset. However, it did cause the overall accuracy to dip from 88% to 84%, and separate tests showed that the OOD detection wrongly flagged 4% of the overall in-distribution dataset as OOD, as shown in Table 3.

*4. Comparison and evaluation*

Although our volume-based Random Forest models performed notably worse than the CN output-based ones, as Table 2 shows, they demonstrated both the usability of image processing to discern essential regions of the brain, as well as the efficacy of using Random Forest models for classification. Training the Random Forest model and predicting the entire test dataset took less than a second. Our CNN model was able to detect more features than manual processing, resulting in higher accuracy. However, in contrast to the volume-based one, the CNN alone took several hours to train, with the CNN-output Random Forest adding minimally to the required computational power. In addition, OOD detection could flag most of the OOD data tested with a comparatively minor effect on the in-distribution data.

## IV. CONCLUSION

More than one in ten people aged 65 and older have Alzheimer's disease (Texas DSHS). The dormant nature of the disease makes people unaware that they have the disease until the effects are detrimental, making early diagnosis crucial. While machine learning has been used on MRI scans to detect Alzheimer's, most of these models require large amounts of processing power and produce false positives when subject to OOD data, which can cause unnecessary treatment and costs. Therefore, we applied OOD detection to our model, enabling it to flag data likely to be OOD. This flagged data is then marked for further review by a physician.

We achieved a 93% detection accuracy and an 87% classification accuracy by using a low-powered Random Forest model on processed images. We then trained a CNN model on our data and utilized the CNN's output to train our Random Forest model, demonstrating the concept of using stacked ensemble models to improve classification in biomedical imaging. This model produced a 98% accuracy for detection and 95% for classification. These findings can be cleanly summarized by Figure 1, which shows the advantages and disadvantages of each model we trained. Afterward, we took the CNN we trained for detection and applied OOD detection. The resulting model could flag OOD images with a 96% accuracy, with a 4% overall accuracy reduction for in-distribution data. This would mean that while most OOD data was correctly flagged, a small percentage of in-distribution was wrongly marked as OOD. This did cause an overall decline in accuracy, but it also allowed our model to avoid the misdiagnosis of OOD samples. Overall, these models were optimized for low processing power with minimized trainable parameters, making them more accessible to physicians in need of a reliable Alzheimer's diagnosis tool.

However, our methodology does come with some limitations. Our dataset was split between "non-demented," "very mild demented," and "mild demented" with a ratio of 67:14:5, and this imbalance could potentially affect the performance of our random forest model. In addition, we had to exclude the moderate dementia class from our model training due to its low number of subjects, so we had a three-class classification problem instead of four. In addition, our OOD detection was optimized for the brain tumor dataset we used, which may cause it to perform sub-optimally for other OOD category images.

Further research concerning this topic would include computing more features from the processed images to train volume-based Random Forests, as Random Forest models have been shown to perform better with more features. In addition, as we excluded the moderate dementia category due to a low number of samples, future work could include this category to improve categorical classification models. In addition, OOD was only tested to flag various brain tumors, and further work could target more conditions that affect brains similarly. This would make it possible for OOD detection to flag more types of anomalies that could be potentially classified as having Alzheimer's.

By reducing the false positives reported by our Alzheimer's detection model and flagging these for professional review, we make machine learning models more reliable for diagnosis. Since OOD data would likely be flagged as such, the models can be more trusted to not produce false diagnoses in these cases and are more trustworthy to be used independently. In addition, our best-performing detection and classification models could perform with comparable accuracy to leading methodologies that require more computational power, making our methodology more accessible to physicians. This takes a step towards the independence of machine learning in diagnosing brain images with human interference only when needed, eliminating a significant disadvantage of using such models to analyze medical images.

## ACKNOWLEDGEMENTS

We thank S. Shailja, Satish Kumar, and Arthur Caetano for their valuable guidance during our research and the writing of this paper. We also thank the University of California, Santa Barbara and the Summer Research Academies program for their support.

## AUTHOR CONTRIBUTION STATEMENT

A.P. thresholded and preprocessed MRI brain scans, researched Alzheimer's statistics, and worked on creating a Grad-CAM. A.P. wrote, proofread, and edited sections of the manuscript and worked on its references. S.P. preprocessed the images, computed brain region volumes, and trained volume-based Random Forest models for detection and classification. S.P implemented OOD onto the CNN model, and wrote manuscript sections concerning volume-based models, as well as OOD detection. M.Z. created the main CNN framework used for OOD, filtered images for training, applied augmentation to images, and created a stacked ensemble model consisting of a CNN and Random Forest model for high accuracy detection and classification. M.Z. wrote sections of the manuscript concerning these areas. All authors reviewed the manuscript.